\title{Improved approximation bounds for Vector Bin Packing}
\author{
        Chetan S. Rao \\
        Department of Computer Science \& Engineering\\
        National Institute of Technology Calicut\\
        Kerala 673601, India
        \and
        Jeffrey John Geevarghese\\
        Department of Computer Science \& Engineering\\
        National Institute of Technology Calicut\\
        Kerala 673601, India
	    \and
	    Karthik Rajan\\
        Department of Computer Science \& Engineering\\
        National Institute of Technology Calicut\\
        Kerala 673601, India
}
\date{\today}

\documentclass[11pt]{article}
\usepackage{amsmath}
\usepackage{algorithm}
\usepackage{algorithmic}
\usepackage{hyperref}

\newcounter{theorem}

\newcommand{\theoremlike}[1]{\par\medskip\penalty-250\refstepcounter{theorem}{\bfseries\scshape\noindent#1 \thetheorem.}\slshape}
\newenvironment{theorem}{\theoremlike{Theorem}}{\par\medskip}
\newenvironment{corollary}{\theoremlike{Corollary}}{\par\medskip}

\newenvironment{lemma}{\theoremlike{Lemma}}{\par\medskip}

{\par\medskip}
{\par\medskip}

\newcommand{\proof}{{\scshape\noindent Proof.\quad}}
\newcommand{\qed}{\hfill\rule{1ex}{1em}\penalty-1000{\par\medskip}}

\begin{document}
\maketitle
\begin{abstract}
In this paper we propose an improved approximation scheme for the Vector Bin Packing problem (\textsc{VBP}), based on the combination of (near-)optimal solution of the Linear Programming (LP) relaxation and a greedy (modified first-fit) heuristic. The Vector Bin Packing problem of higher dimension $(d \ge 2)$ is not known to have asymptotic polynomial-time approximation schemes (unless P = NP).\\
\hspace*{3mm}Our algorithm improves over the previously-known guarantee of $(ln$ $d + 1 + \epsilon)$ by Bansal et al.~\cite{BCS} for higher dimensions $(d>2)$.
We provide a $\theta(1)$ approximation scheme for certain set of inputs for any dimension $d$. More precisely, we provide a 2-OPT algorithm, a result which is irrespective of the number of dimensions $d$.
\end{abstract}

\section{Introduction}
  Packing items of variable sizes into a given space is a fundamental problem of combinatorial optimization. This problem dates back to the origin of Operations Research. The packing problem and it's multidimensional variants have vital practical applications in diverse domains including the cutting, packaging and other industries. In this paper, we consider a type of packing problem called the Vector Bin Packing (Vector Packing) problem and propose a better bound for the multidimensional version of the problem. The items of variable sizes are packed into containers called bins with fixed size along all dimensions.
  
  Multidimensionality plays an important role in capturing incomparable characteristics of the objects that are to be packed. For example, the memory requirements and bandwidth requirements in a distributed computing environment are incomparable. Multidimensionality also suffices the different costs that may be associated with each of these requirements. In this paper, we provide a method which is applicable to any dimension \textit{d}.
  
\subsection{Problem Definition}
We now formulate the optimization problem that we are addressing.\\ \\
\textbf{Vector Bin Packing problem} (VBP)\\
\textit{Given a set S of n d-dimensional vectors $p_{1}$, $p_{2}$, \ldots, $p_{n}$ from [0,1]$^{d}$, find a packing (partition) of S into $A_{1}$, $A_{2}$, \ldots, $A_{m}$ such that $\sum_{p \in A_{i}} p^k \leq 1,$ $\forall i, k$ ($p^{k}$ denotes the projection of vector p along `k'th dimension). The objective is minimize the value of m, the number of partitions.}\\
When $d=1$, the problem is an instance of the classical bin packing problem (BP). 

\subsection{Related work}
One dimensional bin packing problem has been studied extensively. Fernandez
de la Vega and Lueker~\cite{FL} gave the first asymptotic polynomial-time
approximation scheme (APTAS). They put forward a rounding technique that 
allowed them to reduce the problem of packing large items to finding an optimum
packing of just a constant number of items (at a cost of $\epsilon$ times OPT).
Their algorithm was later improved by Karmarkar and Karp~\cite{KK}, to a 
(1+$log^{2}$)-OPT bound.

For 2-dimensional vector bin packing, Woeginger~\cite{W} proved that there is no
APTAS. For higher dimensions, Fernandez de la Vega and Lueker~\cite{FL} proposed a 
simple ($d + \epsilon$)-OPT algorithm, which extends the idea of 1-dimensional bin packing.
Chekuri and Khanna~\cite{CK} showed an $O$(log $d$)-approximation algorithm that
runs in polynomial time for fixed d. Bansal et al.~\cite{BCS} improved this result,
showing an (ln $d$ + 1 + $\epsilon$)-approximation algorithm for any $\epsilon \ge 0$. Karger et 
al.~\cite{KO} have recently proposed a polynomial approximation scheme for randomized instances of 
the multidimensional vector bin packing using smoothing techniques. Patt-Shamir et al.~\cite{BPS} 
have recently explored the vector bin packing problem with bins of varying sizes and propose a 
(ln $2d$ + 1 + $\epsilon$)-approximation algorithm for any $\epsilon \ge 0$.

\subsection{Our results and organization}
In this paper, we provide an improved approximation bound for the vector bin packing problem of any dimension.
We provide a $\theta(1)$ approximation guarantee for certain set of inputs for any dimension $d$. In more specific settings,
we provide a 2-OPT guarantee for large inputs irrespective of the dimension $d$. This is a notable improvement 
over the previously known guarantee of $(ln$ $d + 1 + \epsilon)$ for higher dimensions ($d>2$).

The rest of the paper is organized as follows. Section \ref{probform} gives the linear program (LP) formulation of the problem. In Section \ref{algo}, we provide the approximation algorithm and prove the bounds. 

\section{Problem formulation} \label{probform}
We formulate the problem as an integer program in \ref{ilpform}. The integer constraints are relaxed and we formulate
it's dual (\ref{dualform}). The solution of the relaxed integer program gives a thoughtful insight about the optimal number of bins.

\subsection{Integer Linear Program (ILP) formulation} \label{ilpform}
The vector bin packing problem (\textsc{VBP}) can be formulated as an integer program. We use two binary variables $x_{ij}$ and $y_{j}$. 
The binary variable $x_{ij}$ indicates if vector $p_{i}$ is assigned to bin j and the binary variable $y_{j}$ 
indicates whether bin j is in use or not. Our objective is to minimize the number of bins used.

The number of bins $m$ can initially be set to a sufficiently large value arrived at by any heuristic (example - 
de la Vega and Leuker~\cite{FL} give a $O(d)$-OPT bound on the number of bins). Then, we formulate the integer program 
(\textsc{ILP}) as follows -

\begin{align}
    & \mbox{minimize :\hspace{3mm}} \sum_{j} y_{j} & \\
    \mbox{such that \hspace*{3mm}} & \sum_{j} x_{ij} = 1 & 1 \leq i \leq n \label{lp1} \\ 
    & \sum_{i} p_{i}^{k} . x_{ij} \leq 1 & 1 \leq j \leq m, 1 \leq k \leq d \label{lp2}\\
    & y_{j} \geq x_{ij} & 1 \leq i \leq n, 1 \leq j \leq m\\
    & x_{ij} \in \mbox{\{}0, 1\mbox{\}} & 1 \leq i \leq n, 1 \leq j \leq m
\end{align}

The constraints of the ILP are as follows -
\begin{itemize}
\item × Constraint (2) states that every vector is packed in a bin. 
\item × Constraint (3) ensures that the packed vectors do not exceed the bin dimensions. 
\item × Constraint (4) tells whether a bin is used or not.
\item × Constraint (5) ensures that a vector is either packed entirely in a bin or not. 
\end{itemize}

\paragraph{} Constraint (5) can be relaxed as follows to obtain a linear program (LP).
\begin{equation} \label{lp3}
	x_{ij} \geq 0 \quad \qquad 1 \leq i \leq n,\quad 1 \leq j \leq m  \tag{5a}
\end{equation}
We can obtain a feasible solution for the LP using any standard method~\cite{DA}. Using binary search technique, we can also find the least value of $m$, $m^{'} \in Z^{+}$ for the relaxed ILP for which a feasible solution exists. The value of $m^{'}$ thus obtained will be less than the optimal solution for the integer program i.e. ($m^{'} \leq $ OPT). However, the solution obtained is usually not integral. To tackle this problem, we formulate a dual-maximization problem for the above relaxed ILP.

\begin{theorem} ~\cite{CK} \label{th:1}
Any basic feasible solution to the relaxed LP defined by Equations \ref{lp1}, \ref{lp2}, and \ref{lp3} has at most $d \cdot m$ vectors that are fractionally assigned to more than one bin.
\end{theorem}

\subsection{Dual-maximization problem} \label{dualform}
We introduce several new variables  $z_{ij}$ to formulate the dual. The dual-maximization problem formulation is given in the Appendix \ref{appendix}. We arrive at the following set of equations and constraints -\\
\begin{align}
    & \mbox{maximize :\hspace{3mm}} \sum_{i} \sum_{j} x_{ij}z_{ij} & \\
    \mbox{such that \hspace*{3mm}} & \sum_{j} x_{ij} = 1 & 1 \leq i \leq n\\
    & \sum_{i} p_{i}^{k} . x_{ij} \leq 1 & 1 \leq j \leq m, 1 \leq k \leq d\\
    & \sum_{i} z_{ij} \leq 1 & 1 \leq j \leq m \label{cons9}\\
    & x_{ij}, z_{ij} \geq 0 & 1 \leq i \leq n, 1 \leq j \leq m
\end{align}

This is a nonlinear program (NLP) as the objective function is nonlinear. Hereafter, we shall refer to it as NLP. 

\begin{lemma}
The value of the objective function of the nonlinear program (NLP) lies between 1 and $m$.
\end{lemma}
\proof \\
\textbf{Lower bound} : Let $z_{ij}=\frac{1}{n} \forall(i,j)$. Then, the objective function becomes -
\begin{align*}
	\mbox{Obj = }\sum_{i} \sum_{j} x_{ij} \frac{1}{n} = \frac{1}{n} (\sum_{i} \sum_{j} x_{ij}) = \frac{1}{n} (n) = 1
\end{align*}
\textbf{Upper bound} : The value of the objective function is upper bounded as follows -
\begin{align*}
	\mbox{Obj = }\sum_{i} \sum_{j} x_{ij} z_{ij} \le \sum_{i} \sum_{j} z_{ij} = \sum_{j} (\sum_{i} z_{ij}) \le \sum_{j} (1) = m 
\end{align*} 
Thus, the range of values the objective function will assume is between $1$ and $m$. \qed

The actual range of values that $\sum_{i} \sum_{j} x_{ij}z_{ij}$ (given the constraints) is $0$ to $m$. The objective function gives us a hint 
about the utility of the bins. Intuitively, if the value of the objective function is k, we can assume that the expected utility of each bin
is $\frac{k}{m}$.

\begin{lemma}
In the optimal solution to the nonlinear program (NLP), if any values of $x_{ij}$'s are 0, the corresponding values of $z_{ij}$'s are 0.
\end{lemma}
\proof
	Let $x_{i_{1}j_{1}}$ = 0 for some value of ($i_{1},j_{1}$). From constraint (\ref{cons9}) of the nonlinear program, we have -
	\begin{align*}
		\sum_{i} z_{ij_{1}} = z_{1j_{1}} + z_{2j_{1}} + \ldots + z_{i_{1}j_{1}} + \ldots + z_{nj_{1}}  \le 1
	\end{align*}
	Atleast one vector $i(i \ne i_{1})$ is packed in bin $j_{1}$ (i.e. $x_{ij_{1}} > 0$). For this vector $i$, the value of $z_{ij_{1}} > 0$ in the optimal solution. The term $\sum_{i} x_{ij_{1}} z_{ij_{1}}$ is maximized iff the value of $z_{i_{1}j_{1}}$ is redistributed among other $z_{ij_{1}}$ values where $x_{ij_{1}}>0$. Hence, the value of $z_{i_{1}j_{1}}$ must be 0. \qed

\begin{theorem} \label{th:4}
  Given a sufficiently large value of $m (m \geq OPT)$, the solution (value of $x_{ij}$) returned by the nonlinear program NLP will be integral, given that $z_{ij}$ is non-zero whenever $x_{ij} \ne 0$.
\end{theorem} 

\proof Consider $n$ vectors which are to be packed. Assume that the the optimal packing can be done with 2 bins (i.e. OPT=2). Without loss of generality, let us assume that the vectors $1, 2,\ldots, k_{1}$ $(k_{1} < n)$ be packed in bin 1, and vectors $k_{1}+1, k_{1}+2, \ldots, n$ be packed in bin 2 in the optimal solution OPT. \\
That is, let $x_{11} = x_{21} =\ldots = x_{k_{1}1} = x_{(k_{1}+1)2} = \ldots = x_{n2} = 1$.
The maximum value of the objective function in this case will be -
\begin{align*}
	\mbox{Obj} & = \sum_{i=1}^{k_{1}} z_{i1}  + \sum_{i=k_{1}+1}^{n} z_{i2} \\
				& \le 1 + 1 = 2
\end{align*}
Now, let us assume that the vector $k_{1} + 1$ be fractionally packable in bin 1 and bin 2. Let the fraction that can be packed be $t (0 < t < 1)$ in bin 1 and $(1 - t)$ in bin 2. The objective function in this case can be shown to be -
\begin{align*}
	\mbox{Obj} & = \sum_{i=1}^{k_{1}} z_{i1} + tz_{(k_{1}+1)1} + \sum_{i=k_{1}+2}^{n} z_{i2} + (1-t)z_{(k_{1}+1)2}\\
	 & \leq 1 - (1-t)z_{(k_{1}+1)1} + 1 - tz_{(k_{1}+1)2}\\ 
	 & \leq 2 - (1-t)z_{(k_{1}+1)1} - tz_{(k_{1}+1)2}\\
	 & < 2.
\end{align*}

Similarly, we can extend the same argument when more than 1 vectors from bin 2 are fractionally packed in bin 1. The objective function strictly decreases as follows -
\begin{align*}
	\mbox{Obj} & = \sum_{i=1}^{k_{1}} z_{i1} + \sum_{i=k_{1}}^{n} t_{i} z_{i1} + \sum_{i=k_{1}+1}^{n} (1-t_{i})z_{i2}\\
	 & \leq 1 - \sum_{i=k_{1}+1}^{n} (1-t_{i})z_{i1} + 1 - \sum_{i=k_{1}+1}^{n} t_{i}z_{i2}\\  
	 & \leq 2 - \sum_{i=k_{1}+1}^{n} (1-t_{i})z_{i1} - \sum_{i=k_{1}+1}^{n} t_{i}z_{i2}\\
	 & < 2.
\end{align*}
where $t_{i}$ and (1-$t_{i}$) are the fractional packings of vector $i$ in bin 1 \& 2 respectively.

By induction, the same proof can be extended when the number of required bins is more than 2 (i.e. OPT $>$ 2). Thus, the maximum value of the objective function occurs when the variables $x_{ij}$ are integral. \qed
	
\begin{theorem} \label{th:5}
For the solution to be (near-) optimal and integral, the value of $z_{ij}$ should be a function of $x_{ij}$.
\end{theorem}
\proof The integral part follows from Theorem \ref{th:4}. Let us look into the optimality of the solution obtained.\\
From the Jensen's Inequality, we have that if $f$ is a convex function (``concave-up'') on an interval $I$ and $a_{i} \in I$ then for weights $\lambda_{i}$ summing to 1 -
\begin{align*}
	f(\sum_{i=1}^{n} \lambda_{i} a_{i}) \le \sum_{i=1}^{n} \lambda_{i} f(a_{i})
\end{align*}
We can apply Jensen's inequality with $\lambda_{i}$ and $a_{i}$ corresponding to $x_{ij}$ and $z_{ij}$, respectively. The modified set of equations in this case is as follows -  
\begin{align*}
	f(\sum_{j=1}^{m} x_{ij} z_{ij}) \le \sum_{j=1}^{m} x_{ij} f(z_{ij}) \quad \quad \quad 1 \le i \le n \tag{a} \label{eq:a}
\end{align*}
From the property of convex functions, we have -
\begin{align*}
	f(tx) \le tf(x) \quad \quad \quad 0 \le t \le 1 \tag{b} \label{eq:b}
\end{align*}
From (\ref{eq:a}) and (\ref{eq:b}), we have -
\begin{align*}
	f(\frac{1}{n} \sum_{i=1}^{n} \sum_{j=1}^{m} x_{ij} z_{ij}) \le \frac{1}{n} \sum_{i=1}^{n} \sum_{j=1}^{m} x_{ij} f(z_{ij}) \tag{c} \label{eq:c}
\end{align*}
Since $f(x)$ is a convex function, any value which maximizes $x$ also maximizes $f(x)$ and vice-versa. Hence, from the inequality (\ref{eq:c}), we have that the term $x_{ij} f(z_{ij})$ should be maximized for the objective function to be maximized. Indirectly, $z_{ij}$ has to be maximized relative to the values of $x_{ij}$. The value of $z_{ij}$ is constrained by the constraint (\ref{cons9}), and hence we come up with the following function for $z_{ij}$ -
\begin{align}
	z_{ij} &= \frac{x_{ij}}{\sum_{i} x_{ij}} \label{eq:11} \\
	\implies \sum_{i} z_{ij} &= \sum_{i} \frac{x_{ij}}{\sum_{i} x_{ij}} = 1 \nonumber
\end{align}
Thus, from (\ref{eq:11}), the objective function is as follows -
\begin{align*}
	\mbox{ Obj = } \sum_{i} \sum_{j} x_{ij} z_{ij} = \sum_{i} \sum_{j} x_{ij} (\frac{x_{ij}}{\sum_{i} x_{ij}}) = \sum_{j} \sum_{i} \frac{x_{ij}^2}{\sum_{i} x_{ij}} = \sum_{j} \frac{\sum_{i} x_{ij}^2}{\sum_{i} x_{ij}}
\end{align*}
From the RMS (root mean square), arithmetic mean and standard deviation relation, we have -
\begin{align}
	x_{RMS}^2 &= x_{M}^{2} + \sigma_{x}^{2} \label{eq:12} \\
	\mbox{where \hspace*{3mm} } x_{RMS} &= \sqrt{\frac{x_{1}^{2} + x_{2}^{2} + \ldots + x_{n}^{2}}{n}} \label{eq:13} \\
	x_{M} &= \frac{x_{1} + x_{2} + \ldots + x_{n}}{n} \label{eq:14} \\
	\sigma_{x} &= \sqrt{ \frac{1}{n} \sum_{i=1}^{n} (x_{i} - x_{M})^{2} } \label{eq:15}
\end{align}
On simplifying (\ref{eq:12}), (\ref{eq:13}), (\ref{eq:14}) and (\ref{eq:15}), we have -
\begin{align*}
	\frac{\sum_{i} x_{ij}^2}{\sum_{i} x_{ij}} &= \frac {\sum_{i} x_{ij}}{n} + \frac{n\sigma_{x}^{2}}{\sum_{i} x_{ij}} \\ &= \frac {\sum_{i} x_{ij}}{n} + \frac{n( \frac{1}{n} \sum_{i=1}^{n} (x_{ij} - x_{M})^{2} )}{\sum_{i} x_{ij}}
\end{align*}
From constraint (\ref{lp1}), we have $x_{M} = \frac{1}{m}$ and Theorem \ref{th:1} states that there exists at least $n - d \cdot m$ elements having $x_{ij} = 1$ (i.e. on an average $\frac{n - d \cdot m}{m}$ per bin). \\
Thus,
\begin{align*}
	\frac{\sum_{i} x_{ij}^{2}}{\sum_{i} x_{ij}} &= \frac {\sum_{i} x_{ij}}{n} + \frac{n( \frac{1}{n} \sum_{i=1}^{n} (x_{ij} - \frac{1}{m})^{2} )}{\sum_{i} x_{ij}}\\
				&= \frac {\sum_{i} x_{ij}}{n} + \frac {\sum_{i=1}^{n} ( x_{ij} - \frac{1}{m})^{2} }{\sum_{i} x_{ij}}\\
				&\ge \frac {\sum_{i} x_{ij}}{n} + \frac { \frac{n-dm}{m} ( 1 - \frac{1}{m})^{2} }{\sum_{i} x_{ij}}\\
				&\ge \frac {\sum_{i} x_{ij}}{n} + \frac { \frac{n-dm}{m} ( 1 - \frac{1}{m})^{2} }{\frac{n}{m}} \\
\end{align*}
\begin{align*}
	\implies \sum_{j} \frac{\sum_{i} x_{ij}^{2}}{\sum_{i} x_{ij}} &\ge \sum_{j} \frac {\sum_{i} x_{ij}}{n} + \sum_{j} \frac { \frac{n-dm}{m} ( 1 - \frac{1}{m})^{2} }{\frac{n}{m}} \\
				&= \frac {\sum_{ij} x_{ij}}{n} + m \frac { \frac{n-dm}{m} ( 1 - \frac{1}{m})^{2} }{\frac{n}{m}}\\
				&= \frac {n}{n} + m \frac {(n-dm)( 1 - \frac{1}{m})^{2} }{n}\\
				&\ge 1 + m \frac {(n-dm)( 1 - \frac{2}{m}) }{n}\\
				&= 1 + m (1-\frac{dm}{n})( 1 - \frac{2}{m}) \tag{15a} \label{eq:15a}\\
				&= m - 1 + \frac{dm}{n}(2 - m)\\
				&\approx m - 1 \mbox{\hspace*{5mm} ($n>>dm$)}
\end{align*}
Hence, the value of the objective function ($\approx m-1$) is close to it's upper bound ($= m$) indicating the near optimal usage of the bins. Any basic feasible solution would provide a near optimal packing when $n >> d \cdot m$ \qed

\begin{corollary} \label{cor:mean}
For a fixed value of $\sum_{i} x_{ij}$, the minimum value of $\frac{\sum_{i} x_{ij}^2}{\sum_{i} x_{ij}} $ occurs when all $x_{ij}$'s are equal. Also, the minimum value is equal to $x_{ij}$.
\end{corollary}
\proof From equations (\ref{eq:12}), (\ref{eq:13}), (\ref{eq:14}) and (\ref{eq:15}), we have -
\begin{align*}
	\frac{\sum_{i} x_{ij}^2}{\sum_{i} x_{ij}} &= \frac {\sum_{i} x_{ij}}{n} + \frac{n\sigma_{x}^{2}}{\sum_{i} x_{ij}} \\ 
						  &\ge x_{M}
\end{align*}
The equality holds only when all the $x_{ij}$'s are equal \qed

Any basic feasible solution for the relaxed LP will not necessarily be integral. However, from Theorem \ref{th:5}, we have seen that any such solution will be close to the optimal integer solution. We now present an algorithm which will provide the near optimal integer solution.

\section{Approximation algorithm and it's bounds} \label{algo}
\algsetup{indent=2em}
\begin{algorithm}[h!]
  \caption{$\ensuremath{\mbox{\sc PackingVectors}}(P_{n}, d)$}\label{alg:packvectors}
  \begin{algorithmic}[1]
    \REQUIRE An set of vectors $p_{1}, p_{2}, \ldots, p_{n}$; $P_{n}$.
    \medskip
    \STATE $(m,X) = SolveLP(P_{n}, d)$
    \IF {$m \ge \frac{n}{2}$}
       \RETURN $FirstFit(P_{n}, d)$
    \ELSIF {$m \le \sqrt{\frac{n}{d}}$}
       \RETURN $GreedyLP(P_{n},X,d)$
    \ELSE
	\RETURN $IterativePack(P_{n},X,d)$
    \ENDIF
  \end{algorithmic}
\end{algorithm}

\begin{theorem}
Algorithm 1 provides a $\theta(1)$-optimal guarantee to the Vector Bin Packing problem (VBP).
\end{theorem}
\proof From the solution of the relaxed LP, the packing problem is classified into 3 cases-\\
\textbf{Case 1 :} $m \ge \frac{n}{2}$
The optimal solution for an integer program is a feasible solution for the corresponding relaxed linear program. Thus, the optimal solution for a linear program is lesser than the optimal solution for an integer program (in a minimization problem). Thus, we have that -
\begin{align*}
	m &\le \mbox{OPT} \\
	m &\ge \frac{n}{2}
\end{align*}
Combining the above, we get -
\begin{align*}
	\mbox{OPT} \ge m \ge \frac{n}{2}
\end{align*}
Since the total number of bins cannot exceed the number of vectors $n$, any first fit algorithm in the worst case can 
have an approximation factor of 2.\\ \\
\textbf{Case 2 :} $m \le \sqrt{\frac{n}{d}}$ \\
Since $m \le \sqrt{\frac{n}{d}}$, we have that -
\begin{align*}
	n \ge dm^{2} = m \cdot dm
\end{align*}
Let $n = k \cdot dm$ for some value of $k$. The equation (\ref{eq:15a}) in Theorem \ref{th:5} is then -
\begin{align*}
	Obj &\ge 1 + m (1-\frac{dm}{n})( 1 - \frac{2}{m}) \\
	    &= 1 + m (1-\frac{1}{k})( 1 - \frac{2}{m}) \\
	    &= m -1 - \frac{(m-2)}{k}
\end{align*}
If $k = \Omega(m)$, the equation then becomes -
\begin{align*}
	Obj &= m - 1 - O(1)\\
	    &\ge m - 2 \quad \quad (k \ge m)
\end{align*}
Thus, the value of objective function is $\ge m - 2$, indicating a (near-)optimal utility of each bin. We use this knowledge and apply a greedy heuristic as given in Algorithm 2.

\begin{algorithm}[h!]
  \caption{$\ensuremath{\mbox{\sc GreedyLP}}(P_{n},X,d)$} \label{alg:greedylp}
  \begin{algorithmic}[1]
    \REQUIRE An set of vectors $p_{1}, p_{2}, \ldots, p_{n}$; $P_{n}$ and a set of $x_{ij}$ values $X$.
    \medskip
    \STATE $P_{n}^{'} = P_{n}$
    \STATE $X^{'} = SortDescending(X)$
    \WHILE {$X^{'} \ne \Phi$}
    \STATE Remove the top element $x_{ij}$ in $X^{'}$
    \IF {vector $p_{i}$ fits in bin $j$}
	\STATE $Pack(i,j)$
	\STATE $P_{n}^{'} = P_{n}^{'} \backslash (p_{i})$
    \ENDIF
	\STATE $X^{'} = X^{'} \backslash \{x_{ij}\}$
    \ENDWHILE
    \STATE $PackingVectors(P_{n}^{'}, d)$
  \end{algorithmic}
\end{algorithm}
From Theorem \ref{th:1}, the algorithm completely packs a majority of the vectors into the bins (i.e. $n - dm \ge dm(m-1) \ge \frac{n}{2}$ for $m\ge2$). Less than half of the vectors remain to be packed by the repeated iteration. Hence, the required bins $\le $ 2-OPT.

\textbf{Case 3 :} $\frac{n}{2} \ge m \ge \sqrt{\frac{n}{d}}$ \\
\begin{algorithm}[h!]
  \caption{$\ensuremath{\mbox{\sc IterativePack}}(P_{n},X,d)$}\label{alg:iterativepack}
  \begin{algorithmic}[1]
    \REQUIRE An set of vectors $p_{1}, p_{2}, \ldots, p_{n}$; $P_{n}$ and a set of $x_{ij}$ values $X$.
    \medskip
    \STATE $P_{n}^{'} = P_{n}$
    \STATE $Z = FindDualObj(X, d)$
    \FOR{$j=1$ \TO $m$}
	\IF{$ \sum_{i} x_{ij} z_{ij} \ge \frac{1}{2} $}
		\STATE $X_{j}^{'} = SortDescending(X_{j})$
		\STATE $X_{j}^{''} = RemoveLessThanHalf(X_{j}^{'})$
		\STATE $Pack(X_{j}^{''})$
		\STATE $P_{n}^{'} = P_{n}^{'} \backslash PackedVectors$
	\ENDIF
    \ENDFOR
    \STATE $PackingVectors(P_{n}^{'}, d)$
  \end{algorithmic}
\end{algorithm}
We find all the instances where $ \sum_{i} x_{ij} z_{ij} \ge \frac{1}{2} $. From corollary \ref{cor:mean}, $ \sum_{i} x_{ij} z_{ij} \ge \frac{1}{2} $ happens only when more than half of the non-zero values of $x_{ij}$'s are greater than $\frac{1}{2}$ or the standard deviation is fairly high. Also, the utility of the bin is more than $\frac{1}{2}$.

Only those bins whose utility exceeds half are chosen, and the values of $x_{ij}$'s are sorted in descending order. The values of $x_{ij}$'s less than half are ignored (less than half the number of non-zero $x_{ij}$'s). The value of standard deviation can also be high in which case, we have achieved our objective of increasing the gap between the values of $x_{ij}$.

We now proceed to packing the vectors into their respective bins. Since the number of $x_{ij}$'s thrown away are less than half and the remaining values of $x_{ij}$'s are $\ge \frac{1}{2}$. Atmost twice the number of bins are used to pack these vectors with $x_{ij} \ge \frac{1}{2}$. The remaining vectors are deferred to the next iteration.

Thus, in each stage of the algorithm, the number of bins used is atmost twice the optimal since we are packing vectors with $x_{ij} \ge \frac{1}{2}$. Hence, we have proved that the number of bins used is atmost 2-OPT. \qed

\section{Appendix}

\subsection{Dual formulation of the ILP} \label{appendix}
\begin{align} 
    & \mbox{minimize :\hspace{3mm}} \sum_{j} y_{j} & \\
    \mbox{such that \hspace*{3mm}} & \sum_{j} x_{ij} = 1 & 1 \leq i \leq n \label{1lp1}\\ 
    & \sum_{i} p_{i}^{k} . x_{ij} \leq 1 & 1 \leq j \leq m, 1 \leq k \leq d \label{1lp2}\\
    & y_{j} \geq x_{ij} & 1 \leq i \leq n, 1 \leq j \leq m \label{neq:4}\\
    & x_{ij} \in \mbox{\{}0, 1\mbox{\}} & 1 \leq i \leq n, 1 \leq j \leq m \label{1intgr}
\end{align}
Multiply constraint (\ref{neq:4}) by positive multipliers $z_{ij}$ corresponding to $x_{ij}$'s. Adding all such constraints, we obtain -
\begin{align*}
	(\sum_{i} z_{ij}) y_{j} &\ge \sum_{i} x_{ij} z_{ij} &\quad \quad \quad 1 \leq j \leq m \\
	\sum_{j} (\sum_{i} z_{ij}) y_{j} &\ge \sum_{j} \sum_{i} x_{ij} z_{ij} &
\end{align*}
Further, we have -
\begin{align}
	\sum_{j} y_{j} \ge \sum_{j} (\sum_{i} z_{ij}) y_{j} \ge \sum_{j} \sum_{i} x_{ij} z_{ij} \nonumber\\
	\mbox{ subject to \hspace*{5mm} } \sum_{i} z_{ij} \le 1 \quad \quad \label{1newcons}
\end{align}
Thus, the minimization problem can be reframed as a maximization problem with the constaint (\ref{1newcons}) and objective function being -
\begin{align*}
	\mbox{max : \hspace*{3mm}} \sum_{i} \sum_{j} x_{ij} z_{ij} &
\end{align*}
Adding the new constraints and relaxing constraint (\ref{1intgr}), the dual problem is as follows -
\begin{align*}
    & \mbox{maximize :\hspace{3mm}} \sum_{i} \sum_{j} x_{ij}z_{ij} & \\
    \mbox{such that \hspace*{3mm}} & \sum_{j} x_{ij} = 1 & 1 \leq i \leq n\\
    & \sum_{i} p_{i}^{k} . x_{ij} \leq 1 & 1 \leq j \leq m, 1 \leq k \leq d\\
    & \sum_{i} z_{ij} \leq 1 & 1 \leq j \leq m \\
    & x_{ij}, z_{ij} \geq 0 & 1 \leq i \leq n, 1 \leq j \leq m
\end{align*}
\end{document}